\documentclass[a4paper]{article}

\usepackage[english]{babel}
\usepackage[left=1.in, right=1.in, top=1.in, bottom=1.in]{geometry}
\usepackage[utf8]{inputenc} 
\usepackage[T1]{fontenc}    

\usepackage{hyperref}       
\usepackage{url}            
\usepackage{booktabs}       
\usepackage{amsfonts}       
\usepackage{nicefrac}       
\usepackage{microtype}
\usepackage{lipsum}
\usepackage{amsmath}
\usepackage{graphicx}
\usepackage{mathtools}
\usepackage{booktabs}
\usepackage{multirow}
\usepackage{cite}
\usepackage[ruled,vlined]{algorithm2e}
\usepackage{tikz}
\usepackage{array}
\usepackage{tabularx}

\graphicspath{ {./images/} }

\title{Extending Deep Reinforcement Learning Frameworks in Cryptocurrency Market Making}
\author{
  Jonathan Sadighian \thanks{Completed while associated with SESAMm.  The views presented in this paper are of the author and do not necessarily represent the views of SESAMm.} \\
  SESAMm \\
  \texttt{jonathan.m.sadighian@gmail.com} \\
}
\begin{document}
\maketitle

\begin{abstract}
    There has been a recent surge in interest in the application of artificial intelligence to automated trading. Reinforcement learning has been applied to single- and multi-instrument use cases, such as market making or portfolio management. This paper proposes a new approach to framing cryptocurrency market making as a reinforcement learning challenge by introducing an event-based environment wherein an event is defined as a change in price greater or less than a given threshold, as opposed to by tick or time-based events (e.g., every minute, hour, day, etc.). Two policy-based agents are trained to learn a market making trading strategy using eight days of training data and evaluate their performance using 30 days of testing data. Limit order book data recorded from Bitmex is used to validate this approach, which demonstrates improved profit and stability compared to a time-based approach for both agents when using a simple multi-layer perceptron neural network for function approximation and seven different reward functions. 
\end{abstract}


\section{Introduction} \label{Introduction}
    Applying quantitative methods to market making is a longstanding interest of the quantitative finance community.  Over the past decade, researchers have applied stochastic, statistical and machine learning techniques to automate market making.  These approaches often use limit order book (LOB) data to train a model to make a prediction about future price movements, generally for the purpose of maintaining the optimal inventory and quotes (i.e., posted bid and ask orders at the exchange) for the market maker.  These models are typically trained using the most granular form of event-driven data, level II/III tick data, where an \textit{event} is defined as an incoming order received by the exchange (e.g., new, cancel, or modify).  Alternatively, an event can be defined as a time interval (e.g., every \textit{n} seconds, minutes, hours, etc.).  Every time an event occurs, these models have the opportunity to react (e.g., adjust their quotes), thereby indirectly managing inventory by increasing or decreasing the likelihood of posted order execution.

    Tick-based approaches make assumptions about latency and executions, which may impact the capability of a model to translate simulated results into real-world performance.  Researchers attempt to address this challenge in different ways, such as creating simulation rules around market impact, execution rates, and priority of LOB queues \cite{Zhang_2019, Law2019MarketMU, spooner2018market, morariupatrichi2018statedependent, bacry2013hawkes}.  We propose an alternative approach to address this challenge: Fundamentally change the mindset of strategy creation for automated market making from \textit{latency-sensitive} to \textit{intelligent} strategies using deep reinforcement learning (DRL) with time- and price-based event environments. 

    This paper applies DRL to create an \textit{intelligent} market making strategy, extending the DRL Market Making (DRLMM) framework set forth in our previous work \cite{sadighian2019deep}, which used time-based event environments.  The reinforcement learning framework follows a Markov Decision Process (MDP), where an agent interacts with an environment $E$ over discrete time steps $t$, observes a state space $s_t$, takes an action $a_t$ guided by a policy $\pi$, and receives a reward $r_t$.  The policy $\pi$ is a probability distribution mapping state spaces $s_t \in S$ to action spaces $a_t \in A$.  The agent's interactions with the environment continues until a terminal state is reached.  The return $R_t = \sum^{\infty}_{k=0} \gamma^{k} r_{t+k}$ is the total accumulated return from time step $t$ with discount factor $\gamma \in (0,1]$.  The goal of the agent is to maximize the expected return from each state $s_t$ \cite{Sutton1998}.
    
    Although reinforcement learning has seen many recent successes across various domains \cite{mnih2015humanlevel, SilverHuangEtAl16nature, openai2019dota}, the success of its application in automated trading is highly dependent on the reward function (i.e., feedback signal) \cite{MoodyRLT1998, MoodyRLT1999, MoodyRLT2001}.  Previous research \cite{sadighian2019deep} proposed a framework for deep reinforcement learning as applied to cryptocurrency market making (DRLMM) and demonstrated its capability to generalize across different currency pairs.  The focus of this paper is to extend the research of previous work by evaluating how seven different reward functions impact the agent’s trading strategy, and to introduce a new framework for DRLMM using price-based events.

    This paper is structured as follows: section \ref{Introduction}, introduction; section \ref{RelatedWork}, related research; section \ref{Contributions}, contributions in this paper; section \ref{RewardFuctions}, overview of the seven reward functions; section \ref{EventDrivenEnvironments}, overview of time- and price-based event-driven environments; section \ref{Experiment}, our experiment design and methodology; section \ref{Results} results and analysis; and section \ref{Conclusion}, conclusion and future work.

\section{Related Work} \label{RelatedWork}
    The earliest approaches of applying model-free reinforcement learning to automated trading consist of training an agent to learn a directional single-instrument trading strategy using low-resolution price data and policy methods \cite{MoodyRLT1998, MoodyRLT1999, MoodyRLT2001, Gold2003FXTV}.  These approaches find risk-based reward functions, such as Downside Deviation Ratio or Differential Sharpe Ratio, generate more stable out-of-sample results than using actual profit and loss. 

    More recently, researchers have applied reinforcement learning methods to market making.  \cite{spooner2018market} created a framework using value-based model-free methods with high-resolution LOB data from equity markets.  Under this framework, an agent takes a step through the environment every time a new tick event occurs.  They proposed a novel reward function, which dampens the change in unrealized profit and loss asymmetrically, discouraging the agent from speculation.  Although their agent demonstrates stable out-of-sample results, assumptions about latency and executions in the simulator make it unclear how effectively the trained agent would perform in live trading environments.  \cite{patel2018optimizing} proposed a hierarchical reinforcement learning architecture, where a \textit{macro} agent views low-resolution data and generates trade signals, and a \textit{micro} agent accesses the LOB and is responsible for order execution.  Under this approach, the macro agent takes a step through the environment with time-based events using one-minute time intervals; the micro agent interacts with its environment in ten-second intervals.  Although the agent outperformed their baseline, the lack of inventory constraints on the agent makes the results uncertain for live trading. 

    Previous work \cite{sadighian2019deep} proposed a new framework for applying deep reinforcement learning to cryptocurrency market making.  The approach consists of using a time-based event approach with one-second snapshots of LOB data (including derived statistics from order and trade flow imbalances and indicators) to train policy-based model-free actor-critic algorithms.  The performance of two reward functions were compared on Bitcoin, Ether and Litecoin data from Coinbase exchange.  The framework’s ability to generalize was demonstrated by applying trained agents to make markets in different currency pairs profitably. This paper extends previous work through comparing five additional reward functions, and introduce a new approach for the DRLMM framework using price-based events.

\section{Contributions} \label{Contributions}
    The main contributions of this paper are as follows: 
    \begin{enumerate}
    
      \item \textbf{Analysis of seven reward functions} : We extend previous work, apply the DRLMM framework to more reward functions and evaluate the impact on the agent’s market making strategy.  The reward function definitions are explained in section \ref{RewardFuctions}.
      
      \item \textbf{Price-based event environment} : We propose a new approach to defining an event in the agent’s environment and compare this approach to our original time-based environment framework.  The price-based event approach is explained in section \ref{EventDrivenEnvironments}.
      
    \end{enumerate}

\section{Reward Functions} \label{RewardFuctions}
    The reward function serves as a feedback signal to the agent and therefore directly impacts the agent’s trading strategy in a significant way. There are seven reward functions described in this section, which are categorized as profit-and-loss (PnL), goal-oriented, and risk-based approaches.  These seven reward functions provide a wide range of feedback signals to the agent, from frequent to sparse.
    
    When calculating realized PnL, orders are netted in FIFO order and presented in percentage terms, opposed to dollar value, to ensure compatibility if applied to different instruments (all simulation rules are set forth in section \ref{BusinessRules}).

    \subsection{PnL-based Rewards}

        \paragraph{Unrealized PnL} \label{Reward:default}
            The agent's unrealized PnL $UPnL$ provides the agent with a continuous feedback signal (assuming the agent is trading actively and maintains inventory).  This reward function is calculated by multiplying the agent's inventory count $Inv$ by the percentage change in midpoint price $\Delta m$ for time step $t$.  Note, inventory count $Inv$ is an integer, because the agent trades with equally sized orders (see section \ref{BusinessRules} for the comprehensive list of trading rules in our environment).
        
            \begin{equation}
                UPnL_{t} = Inv_t \Delta m_t
            \end{equation}
            where $\Delta m  = \frac{m_t}{m_{t-1}} - 1$ and $Inv_t = \sum_{n=0}^{IM} Ex_t^{n}$ is the total count of executed orders $Ex$ held in inventory and $IM$ is the maximum permitted inventory.  In our experiment, we set $IM = 10$, meaning the agent can execute and hold 10 trades (of equal quantity).

        \paragraph{Unrealized PnL with Realized Fills} \label{Reward:defaultWithFills}
            The unrealized PnL with realized fills $UPnLwF$ reward function (referred to as \textit{positional PnL} in our previous work) is similar to $UPnL$, but includes any realized gains or losses $RPnL^{step}$ obtained between time steps $t$ and $t-1$.  This reward function provides the agent with a continuous feedback signal, as well as larger sparse rewards (assuming the agent is trading actively and maintains inventory).
        
            \begin{equation}
                UPnLwF_{t} = UPnL_t + RPnL^{step}_t
            \end{equation}
            where $RPnL_t^{step} = \left[ \frac{Ex^{E, short}_t}{Ex^{X, cover}_t} - 1 \right] + \left[ \frac{Ex^{X, sell}_t}{Ex^{E, long}_t} - 1 \right]$ and $Ex^{E, long, short}_t$ is the average entry price and $Ex^{X, sell, cover}$ is the average exit price of the executed order(s) between time steps $t$ and $t-1$ for $long$ or $short$ sides.

        \paragraph{Asymmetrical Unrealized PnL with Realized Fills} \label{Reward:asymmetrical}
            The asymmetrical unrealized PnL with realized fills $Asym$ reward function is similar to $UPnLwF$, but removes any upside unrealized PnL to discourage price speculation and adds a small rebate (i.e., half the spread) whenever an open order is executed to promote the use of limit orders.  This reward function is provides both immediate and sparse feedback to the agent (assuming the agent is trading actively and maintains inventory).  Our implementation is similar to \cite{spooner2018market}'s asymmetrically dampened PnL function, but includes the realized gains $RPnL^{step}_t$ from the current time step $t$, which improved our agent's performance in volatile cryptocurrency markets.
            
            \begin{equation}
                Asym_{t} = min(0, \eta UPnL_t) + RPnL^{step}_t + \psi_t
            \end{equation}
            where $\psi = Ex^{n}_t [\frac{m_t}{p^{bid}_t} - 1]$ is the number $n$ of matched (i.e., executed) orders $Ex$ multiplied by half the spread $\frac{m_t}{p^{bid}_t} - 1$ in percentage terms, and $\eta$ is a constant value used for dampening.  In our experiment, we set $\eta$ to 0.35.

        \paragraph{Asymmetrical Unrealized PnL with Realized Fills and Ceiling} \label{Reward:TCA}
            The asymmetrical unrealized PnL with realized fills and gains ceiling $AsymC$ reward function can be though of as an extension of $Asym$, where we add a cap $\kappa$ on the realized upside gains $RPnL^{step}_t$ and remove the half-spread rebate $\psi_t$ on executed limit orders.  The intended effect is that $AsymC$ is discouraged from long inventory holding periods and price speculation due to the ceiling and asymmetrical dampening. Like $Asym$, this reward function provides both immediate and sparse feedback to the agent.
        
            \begin{equation}
                AsymC_{t} = min(0, \eta UPnL_t) + min(RPnL^{step}_t, \kappa)
            \end{equation}
    	    where $\kappa$ is the effective ceiling on time step realized gains. In our experiment, we set $\kappa$ to twice the market order transaction fee.

        \paragraph{Realized PnL Change} \label{Reward:RPNL}
            The change in realized PnL $\Delta RPnL$ provides the agent with a sparse feedback signal since values are only generated at the end of a round-trip trade.  The reward is calculated by taking the difference in realized PnL $RPnL$ values between time step $t$ and $t-1$.
        
            \begin{equation}
                \Delta RPnL_{t} = RPnL_{t} - RPnL_{t-1}
            \end{equation}
            where $RPnL$ is the agent's realized PnL at time step $t$ and previous time step $t-1$.
    
    \subsection{Goal-based Rewards}
        
        \paragraph{Trade Completion} \label{Reward:TC}
        	The trade completion $TC$ reward function provides a goal-oriented feedback signal, where a reward $r_t \in [-1, 1]$ is generated if a objective is obtained or missed.  Moreover, if the realized PnL $RPnL^{step}$ is greater (or less) than a predefined threshold $\varpi$, the reward $r_t$ is 1 (or -1) otherwise, if $RPnL^{step}$ is in between the thresholds, the actual realized PnL in percentage terms is the reward.  Using this approach, the agent is encouraged to open and close positions with a targeted profit-to-loss ratio, and is not rewarded for longer term price speculation.
    		
    		\begin{equation}
        	    TC_{t} = 
        		\begin{cases}
        			1, & \textnormal{if } RPnL^{step}_t \geq \epsilon \varpi \\
        			-1, & \textnormal{if } RPnL^{step}_t \leq - \varpi \\
        			RPnL^{step}_t, & \textnormal{otherwise} \\
        		\end{cases}
        	\end{equation}
        	where $\epsilon$ is a constant used for the multiplier and $\varpi$ is a constant used for the threshold.  In our experiment, we set $\epsilon$ to 2 and $\varpi$ to the market order transaction fee.

    \subsection{Risk-based Rewards}
        \paragraph{Differential Sharpe Ratio} \label{Reward:DSR}
            The differential sharpe ratio $DSR$ provides the agent with a risk adjusted continuous feedback signal (assuming the agent is trading actively and maintains inventory).  Originally proposed \cite{MoodyRLT1998} more than 20 years ago, this reward function is the online version of the well known Sharpe Ratio, but can be calculated cheaply with O(1) time complexity, thereby making it the more practical choice for training agents using high-resolution data sets.
        
            \begin{equation}
                DSR_t = \frac{ B_{t-1} \Delta A_t - \frac{1}{2} A_{t-1} \Delta B_t}{( B_{t-1} - A_{t-1}^2)^{3/2}}
            \end{equation}
            where $A_t = A_{t-1} + \eta (R_t - A_{t-1})$ and $B_t = B_{t-1} + \eta (R^2_t - B_{t-1})$ and $\Delta A = R - A_{t-1}$ and $\Delta B = R^2 - B_{t-1}$ and $\eta$ is a constant value. In our experiment, we use $UPnL$ (as described in section \ref{Reward:default}) for $R$ and set $\eta$ to 0.01.

\section{Event-driven Environments} \label{EventDrivenEnvironments}
    When applying reinforcement learning to financial time series, the typical approach to framing the MDP is to have the agent take a step through the environment using a time-based interval. Depending on the trading strategy, the interval of time can be anywhere from seconds to days. For market making, the typical approach is to use \textit{tick} events (e.g., new, cancel or modify order) as the catalyst for an agent to interact with its environment. The tick-based approach differs from a time-based approach in that the events are irregularly spaced in time, and occur in much greater frequency (more than a magnitude). Although tick-based strategies can yield very impressive results in research, external factors (e.g., partial executions, latency, risk checks, etc.) could limit their practicality in live trading environments. We address the challenge by proposing the use of \textit{price}-based events for market making trading strategies, which partially removes the dependency on these assumptions, enabling the deep reinforcement algorithms to learn non-linear market dynamics across multiple time steps (i.e., not latency-sensitive).

    \subsection{Time-based Events} \label{TimeEvents}
        The time-based approach to event-driven environments consists of sampling the data at periodic intervals evenly spaced in time (e.g., every second, minute, day, etc.).  This approach is the most intuitive for trading strategies, since market data is easily available in this format.  This experiment takes snapshots of the LOB (and other inputs in our feature space) using one-second time intervals to reduce the number of events in one trading day from millions to 86,400 (the number of seconds in a 24-hour trading day), resulting in less clock time required to train our agent.
    
    \subsection{Price-based Events} \label{PriceEvents}
        The price-based approach to event-driven environments consists of an \textit{event} being defined as a change in midpoint price $m$ greater or less than a threshold $\beta$.  Following this approach, our data set is further down-sampled from its original form of one-second time intervals into significantly fewer price change events that are irregularly spaced in time, thereby decreasing the amount of time required to train the agent per episode (i.e., one trading day).  In this experiment, the minimum threshold $\beta$ is set to one basis point (i.e., 0.01\%) and use the one-second LOB snapshot data (as described in section \ref{TimeEvents}) as the underlying data set.
        
        \begin{algorithm}[H]
            \SetAlgoLined
            \KwResult{Observation and accumulated reward at time t+n}
            
            $\beta \gets 0.01\%$ \\
            
            $n \gets 0$ \\
            
            $m_t \gets \frac{p^{ask}_t + p^{bid}_t}{2}$ \\
            
            $upper \gets m_t  (1 + \beta)$ \\
            
            $lower \gets m_t  (1 - \beta)$ \\
            
            $step \gets True$ \\
            
            \While{step}{
                \eIf{$upper \le m_{t+n} \le lower$}{
                    $n \gets n + 1$ \\
                }{
                    $step \gets False$ \\ 
                }
            }
            \caption{Deriving price-based events from high-resolution data sets.}
            \label{Alg:PriceEvent}
        \end{algorithm}

\section{Experiment} \label{Experiment}
    In this section, the design and methodology aspects of the experiment are set forth. 
    
    \subsection{Environment Design} \label{EnvironmentDesign}
        
        \subsubsection{Observation Space}
            The agent’s observation space is represented by a combination of LOB data from the first 20 rows, order and trade flow imbalances, indicators, and other hand-crafted indicators.  For each observation, we include 100 window lags.  The observation space implementation specifications are detailed in the appendix (section \ref{AgentObservationSpace}).
        	
        	It is worth noting that in previous work \cite{sadighian2019deep}, the non-stationary feature \textit{price level distances to midpoint} is included in the agent’s observation space; however, this feature does not inform the agent when using Bitmex data.  This is likely due to the tick size at Bitmex being relatively large (0.50) compared to Coinbase exchange (0.01).  As a result, the distances of price levels to the midpoint remain unchanged for 99.99\% of the time at Bitmex.
        
        \subsubsection{Action Space}
        	The agent action space consists of 17 possible actions.  The idea is that the agent can take four general actions: no action, symmetrically quote prices, asymmetrically skew quoted prices, or flatten the entire inventory.  The action space is outlined in Table \ref{table:action_space}.
        	
        	\begin{table}
        		\centering	
        		\begin{tabular}{lccccccccccccccc}
        			\toprule\toprule
        			Action ID & 2 & 3 & 4 & 5 & 6 & 7 & 8 & 9 & 10 & 11 & 12 & 13 & 14 & 15 & 16 \\
        			\midrule
        			Bid & 0 & 0 & 0 & 4 & 4 & 4 & 4 & 9 & 9 & 9 & 9 & 14 & 14 & 14 & 14 \\
        			Ask & 4 & 9 & 14 & 0 & 4 & 9 & 14 & 0 & 4 & 9 & 14 & 0 & 4 & 9 & 14 \\
        			\midrule
        			Action 1 & \multicolumn{15}{l}{No action} \\
        			Action 17 & \multicolumn{15}{l}{Market order $M$ with size $Inv$} \\
        			\bottomrule
        		\end{tabular}
        		\caption{The agent action space with 17 possible actions.  The numbers in the $Bid$ and $Ask$ rows represent the price level at which the agent's orders are set to for a given action and are indexed at zero.  For example, action 2 indicates the agent open orders are skewed so that its bid is at level zero (i.e., best bid) and its ask is at level five.}
        		\label{table:action_space}
        	\end{table}
    
        \subsubsection{Function Approximator}
        	The function approximator is a multilayer perceptron (MLP), which is a forward feed artificial neural network.  The architecture of our implementation consists of 3-layer network with a single shared layer for feature extraction, followed by separate actor and critic networks.  ReLu activations are used in every hidden layer.
    
            \begin{figure}[h]
                \hfill \includegraphics[scale=.15]{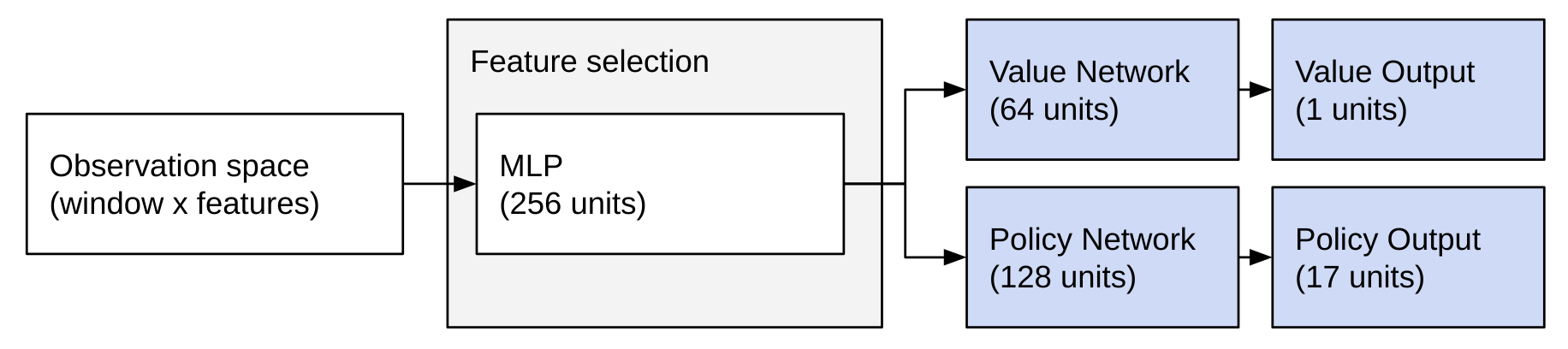}  \hspace*{\fill}
                \caption{Architecture of actor-critic MLP neural network used in the experiments. \emph{Gray} represents shared layers. \emph{Blue} represents non-shared layers.  The window size $w$ is 100 and the feature count varies depending on the feature set, as described in table \ref{table:FeatureCombinations}.}
                \label{fig:MLP_Architecture}
            \end{figure}

        \subsubsection{Reward Functions}
    	    We implement seven different reward functions, as outlined in section \ref{RewardFuctions}.

    \subsection{Agents}
        Two advanced policy-based model-free algorithms are used as market making agents: Advantage Actor-Critic (A2C) and Proximal Policy Optimization (PPO).  We use the Stable Baselines \cite{stable-baselines} implementation for the algorithms.  Since both algorithms run on multiple processes, they require nearly the same amount of clock time to train.  The same policy network architectures (figure \ref{fig:MLP_Architecture}) are used across all experiments, and parameter settings are listed in the appendix (section \ref{AgentConfigurations}).
        	
        	\subsubsection{Advantage Actor-Critic (A2C)}
            	The A2C is an on-policy model-free actor-critic algorithm that is part of policy-based class of RL algorithms.  It interacts with the environment asynchronously while maintaining a policy $\pi (a_{t}|s_{t}; \theta)$ and estimate of the value function $V(s_{t}; \theta_{v})$, and synchronously updates parameters using a GPU, opposed to its off-policy asynchronous update counterpart A3C \cite{mnih2016asynchronous}.  A2C learns good and bad actions through calculating the advantage $A(s_{t} | a_{t})$ of a particular action for a given state.  The advantage is the difference between the action-value $Q(s_{t} | a_{t})$ and state value $V(s_{t})$.  The A2C algorithm also uses k-step returns to update both policy and value-function, which results in more stable learning than a vanilla policy gradient, which uses 1-step returns.  These features, asynchronous training and k-step returns, make A2C a strong fit for its application to market making, which relies on noisy high-resolution LOB data.
            	
            	The A2C update is calculated as	
            	\begin{equation}
            		\nabla_{\theta} J(\theta) = \nabla_{\theta '} log \pi (a_{t} | s_{t}; \theta ') A(s_{t}, a_{t}; \theta, \theta_{\upsilon})
            	\end{equation}
            	
            	where $A(s_{t}, a_{t}; \theta, \theta_{\upsilon})$ is the estimate of the advantage function given by $\sum_{i=0}^{k-1} = \gamma^{i} r_{t + i}  + \gamma^{k} V(s_{t + k}, \theta_{\upsilon}) - V(s_{t}, \theta_{\upsilon})$, where $k$ can vary by an upper bound $t_{max}$ \cite{mnih2016asynchronous}.
        	
        	\subsubsection{Proximal Policy Optimization (PPO)}
            	The PPO is an on-policy model-free actor-critic algorithm is part of policy-based class of RL algorithms, even though the policy is indirectly updated through a surrogate function.  Like the A2C algorithm, it interacts with the environment asynchronously, makes synchronous parameter updates $\theta$, and uses k-step returns. However, unlike A2C, PPO uses Generalized Advantage Estimation (GAE) to reduce the bias of advantages \cite{Schulman2015HighDimensionalCC} and indirectly optimizes the policy $\pi_{\theta}(a_{t} | s_{t})$ through a clipped surrogate function $L^{CLIP}$ that represents the difference between the new policy after the most recent update $\pi_{\theta}(a | s)$ and the old policy before the most recent update $\pi_{\theta_{k}}(a | s)$ \cite{schulman2017proximal}.  This surrogate function removes the incentive for a new policy to depart from the old policy, thereby increasing learning stability.  These features, asynchronous training, k-step returns, and surrogate function, make PPO a strong fit for its application to market making, which relies on noisy high-resolution LOB data.
            	
            	The PPO Clip update is calculated as		
            	\begin{equation}
            		\small L(s,a,\theta_k,\theta) = \min\left( \frac{\pi_{\theta}(a_{t}|s_{t})}{\pi_{\theta_k}(a_{t}|s_{t})} A^{\pi_{\theta_k}}(s_{t},a_{t}), \;\;
            		\textnormal{clip}\left(\frac{\pi_{\theta}(a_{t}|s_{t})}{\pi_{\theta_k}(a_{t}|s_{t})}, 1 - \epsilon, 1+\epsilon \right) A^{\pi_{\theta_k}}(s_{t},a_{t}) \right)
            	\end{equation}
            	
            	\normalsize where $\epsilon$ is a hyperparameter constant \cite{schulman2017proximal}.

    \subsection{Methodology} \label{Methodology}
    
        \subsubsection{Data Collection} \label{DataCollection}
            LOB data for cryptocurrencies is free to access via WebSocket, but not readily downloadable from exchanges, and therefore requires recording. The data set for this experiment was recorded using Level II tick and trade data from Bitmex exchange\footnote{https://www.bitmex.com/} and persisted into an Arctic TickStore\footnote{https://github.com/man-group/arctic} for storage. 

            Unlike previous work, where we replayed recorded data to reconstruct the data set, in this experiment we recorded the LOB snapshots in real-time using one-second time intervals. This approach has two main advantages over replaying tick data.  First, the computational burden is significantly reduced (millions of tick events per trading day) in setting up the experiment, since the LOB no longer needs to be reconstructed to create LOB snapshot data.  Second, the data feed is more reflective of a production trading system.  However, this approach introduced a small amount of latency into the snapshot intervals (less than 1 millisecond), resulting in approximately 86,390 snapshots per a 24 hour trading day, opposed to the actual number of seconds (86,400).  We export the recorded data to compressed CSV files, segregated by trading date using UTC timezone; each file is approximately 160Mb in size before compression. 
        
        \subsubsection{Data Processing} \label{DataProcessing}
            Since LOB data cannot be used for machine learning without preprocessing, it is necessary to apply a normalization technique to the raw data set.  In this experiment, the data set was normalized using the approach described by \cite{tsantekidis2018, sadighian2019deep}, which transforms the LOB from non-stationary into a stationary feature set, then uses the previous three trading days to fit and $z$-score normalize the current trading day’s values, and in which data point $z_x$ is $\sigma$ standard deviations from the mean $\bar{x}$.  After normalizing the data set, outliers (values less than -10 or greater than 10) are clipped. 
        	
        	\begin{equation}
        	    z_{x} = \frac{x - \bar{x}}{\sigma}
        	\end{equation}
        
        \subsubsection{Simulation Rules} \label{BusinessRules}
            The environment follows a set of rules to ensure the simulation as realistic as possible.
            
            \paragraph{Episode}
             An episode is defined as a 24-hour trading day, using coordinated universal time (UTC) to segregate trading days.  At the end of an episode, the agent is required to flatten its entire inventory as a risk control. 
            
            \paragraph{Transaction Fees}
            Transaction fees for orders are included and are deducted from the realized profit and loss when an order is completed.  We use a maker rebate of 0.025\% and taker fee of 0.075\%, which corresponds to Bitmex’s fee schedule at the time of the experiment.  The maker-taker fee structure is crucial to the success of our agent’s market making trading strategy.
            
            \paragraph{Risk Limits}
            The agent is permitted to open one order per side (e.g., bid and ask) at a given moment, and can hold up to ten executed orders (i.e., inventory maximum $IM = 10$) in its inventory $Inv$.  All orders placed by the agent are equal in size $Sz$.  There are no stop losses imposed on agents.
            
            \paragraph{Position Netting}
            If the agent has an open long (short) position and fills an open short (long) order, the existing long (short) position is netted, and the position's realized profit and loss is calculated in FIFO order.  PnL is calculated in percentage terms.
            
            \paragraph{Executions}
            Each time a new order is opened by the agent, the dollar value (e.g., price $\times$ quantity) of the order's price-level $i$ at the time step $t$ is captured by our simulator, and only reduced when there are buy (sell) transactions at or above (below) the ask (bid).  Only after the LOB price-level queue is depleted, can the agent's order begin to be executed.  This environment rule is necessary to help simulate more realistic results.  Additionally, the agent can modify an existing open order, even if it is partially filled, to a new price and reset its priority in the price-level queue.  Once the order is filled completely, the average execution price $Ex^{Avg}$ is used for profit and loss calculations and the agent must wait until the next environment \textit{step} to select an action $a_t$ (such as replenishing the filled order in the order book).
            
            \paragraph{Slippage}
            If the agent selects decides to flatten its inventory, we account for market impact by applying a fixed slippage percentage $\xi$ to each transaction $n$ individually and recursively (e.g., $p^{slippage}_{n} = p_{n-1}^{slippage} \pm \xi$), where $\xi$ is 0.01\% and $\pm$ is linked to order direction.  We noticed adding slippage to the \textit{flatten all} action encouraged the agent to use limit orders more frequently.
        
        \subsubsection{Training and Testing} \label{TrainingAndTesting}
            The market-making agents (A2C and PPO) are trained on 8 days of data (December 27th, 2019 to January 3rd, 2020) and tested on 30 days of data (January 4th to February 3rd, 2020) using perpetual Bitcoin data (instrument: XBTUSD).  Each trading day consists of $\approx$86,390 snapshots, which reflects roughly one snapshot per second in a 24-hour market (slightly less due to latency in our Python implementation, as noted in section \ref{DataCollection}).  Each agent’s performance is evaluated on daily and cumulative returns. 
            
            In each experiment, agents are trained for one million environment steps and episodes restart at a random step in the environment to prevent deterministic learning.  The time-based environment takes advantage of action repeats, enabling agents to accelerate learning; we use five action repeats in our experiment, which results in up to approximately 17,000 agent interactions with the environment per episode.  The price-based environment does not use action repeats, since the number of interactions with the environment is already reduced to approximately 5,000 instances per episode.  It is important to note that during action repeats or between price events, the agent’s action is only performed once and not repeated; all subsequent repeats in the environment consist of taking “no action,” thereby avoid performing illogical repetitive actions multiple times in a row, such as flattening the entire inventory, or re-posting orders to the same price and losing LOB queue priority.  All the experiment parameters are outlined in the appendix (section \ref{AgentConfigurations}).
        
            \begin{table}[]
                \centering
                \begin{tabular}{lcccc}
			        \toprule\toprule
                    \multicolumn{5}{c}{\textbf{Feature Sets}} \\ \hline
                    \multicolumn{1}{l|}{\textit{\textbf{Combination}}} & \multicolumn{1}{l}{\textit{\textbf{LOB Quantity}}} & \multicolumn{1}{r}{\textit{\textbf{Order Flow}}} & \multicolumn{1}{l}{\textit{\textbf{LOB Imbalances}}} & \multicolumn{1}{l}{\textit{\textbf{Indicators}}} \\ \hline
                    \multicolumn{1}{l|}{\textit{Set 1}} & \checkmark & \checkmark & \checkmark & \checkmark \\ \hline
                    \multicolumn{1}{l|}{\textit{Set 2}} &  &  & \checkmark & \checkmark \\ \hline
                    \multicolumn{1}{l|}{\textit{Set 3}} & \checkmark &  & \checkmark & \checkmark \\ \hline
                    \multicolumn{1}{l|}{\textit{Set 4}} & \checkmark &  &  & \checkmark \\ \hline
                    \multicolumn{1}{l|}{\textit{Set 5}} & \checkmark & \checkmark &  & \checkmark \\ \hline
                    \multicolumn{1}{l|}{\textit{Set 6}} &  & \checkmark &  & \checkmark \\  \bottomrule
                \end{tabular}
    		\caption{Combination of features which make up the observation space in different experiments.  For example, \textit{Set 1} uses all available features, whereas \textit{Set 2} uses only LOB Imbalances and indicators to represent the environment's state space.  Implementation details for features are outlined in section \ref{ess}.}
    		\label{table:FeatureCombinations}
            \end{table}

\section{Results} \label{Results}
    In this section, the performance of each agent (PPO and A2C) are compared using the cumulative return (in percentage) including transaction costs from the out-of-sample tests. Our benchmark is a simple buy-and-hold strategy, where we assume Bitcoin is purchased on the first day of the out-of-sample data and sold on the last day, for a total holding period of 30\footnote{\label{DroppedConnection} Not including January 14th, 2020 due to a dropped WebSocket connection} consecutive trading days. Although the train-test split of data sets was selected based on data availability and not empirically, the out-of-sample data set coincidentally captures a volatile upward month long trend in January 2020, which enables the benchmark to generate a 16.25\%\footnotemark[\value{footnote}] return during this period. 

    The best result obtained from our agents is a 17.61\%\footnotemark[\value{footnote}] return over this same period, using reward function Trade Completion $TC$, A2C algorithm, and feature combination \textit{Set 3} for the observation space.  Although the A2C algorithm outperformed the PPO agent in terms of greatest return and number of profitable experiments, it is interesting that no clear trends emerged for the best observation space combination, or reward function (other than what does not work). 

    It is worth noting that on January 19, 2020, the price of Bitcoin sold off more than 5\% in less than 200 seconds, and all experiments (agents, reward functions, and observation space combinations) incurred significant losses ranging between 5\% and 10\% as a result of the rapid price drop; if this trading day were excluded, many more experiments would have yielded positive results.  All experiment results are outlined in tables \ref{table:TimeEvent} and \ref{table:PriceEvent}.

    \subsection{Reward Functions} \label{Results:RewardFunctions}
        We evaluated seven different reward functions across a combination of features in the observation space, A2C and PPO reinforcement learning algorithms, and time- and price-based event environments.  Each reward function resulted in the agent learning a different approach to trading and maintaining its inventory.
        
        \subsubsection{PnL-based rewards}
            Reward functions where realized gains are not incorporated in the function’s feedback signal tended to result in nearsighted trading behavior.  For example, the unrealized profit-and-loss function $UPnL$ encouraged the agent to use market orders (e.g., action 17 - flatten inventory) often and executed many trades with short holding periods, resulting in consistent losses due to transaction costs. 
        
            Reward functions where the feedback signal is sparse tended to result in speculative trading behavior. For example, the change in realized profit function $\Delta RPnL$ encouraged the agent to hold positions for an extended period, regardless if the agent had large unrealized gains or drawdown. 

            Reward functions where the feedback signal is dampened asymmetrically tended to result in tactical trading while failing to exploit large price movements.  For example, the asymmetrical unrealized PnL with realized fills function $Asym$ discouraged the agent holding a position for an extended period of time into a price jump, either resulting in closing out a position too early and foregoing profits, or closing out a position during a transitory drawdown period.  These types of reward functions are very sensitive to the dampening factor $\eta$, and the value 0.35 yielded the most stable out-of-sample performance through a grid search.
        
        \subsubsection{Goal-oriented rewards}
            The Trade Completion $TC$ reward function tended to result in more active trading and inventory management.  For example, agent does not hold positions for speculation and quickly closes positions as they approach the upper and lower boundaries of the reward function curve.
            
        \subsubsection{Risk-based rewards}
            The Differential Sharpe Ratio $DSR$ reward function produced inconsistent results, and appears to be very sensitive to experiment settings.  For example, in some experiments the agents learned very stable trading strategies, while unable to learn at all in other experiments (even with different random seeds).  Additionally, in certain market conditions the agents were able to learn how to exploit price jumps, while making nonsensical decisions in other market regimes.  It is possible that this reward function could have better performance with a thorough parameter grid search.
        
        \begin{figure}[!ht]
            \hfill 
            \includegraphics[scale=.22]{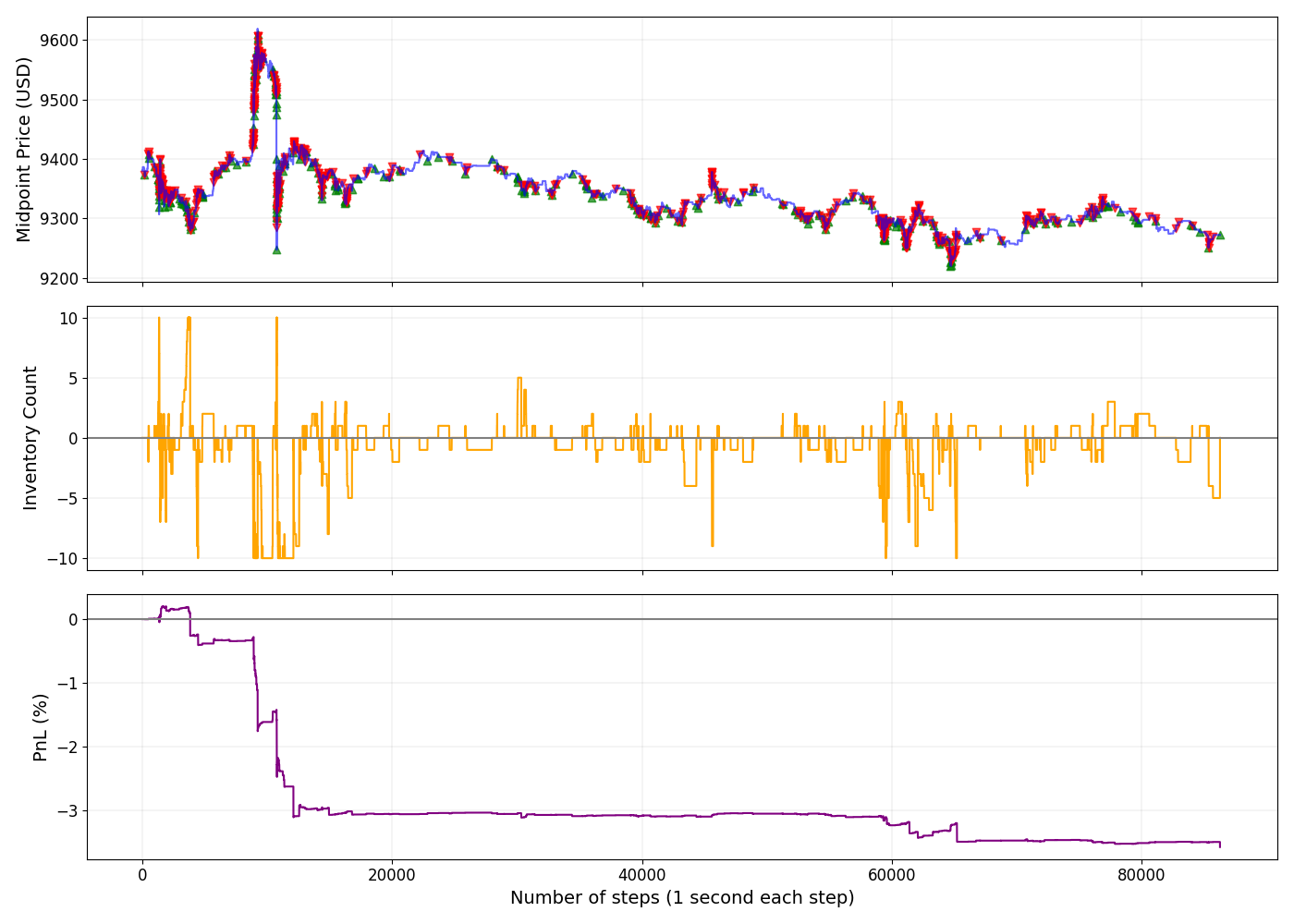}
            \includegraphics[scale=.22]{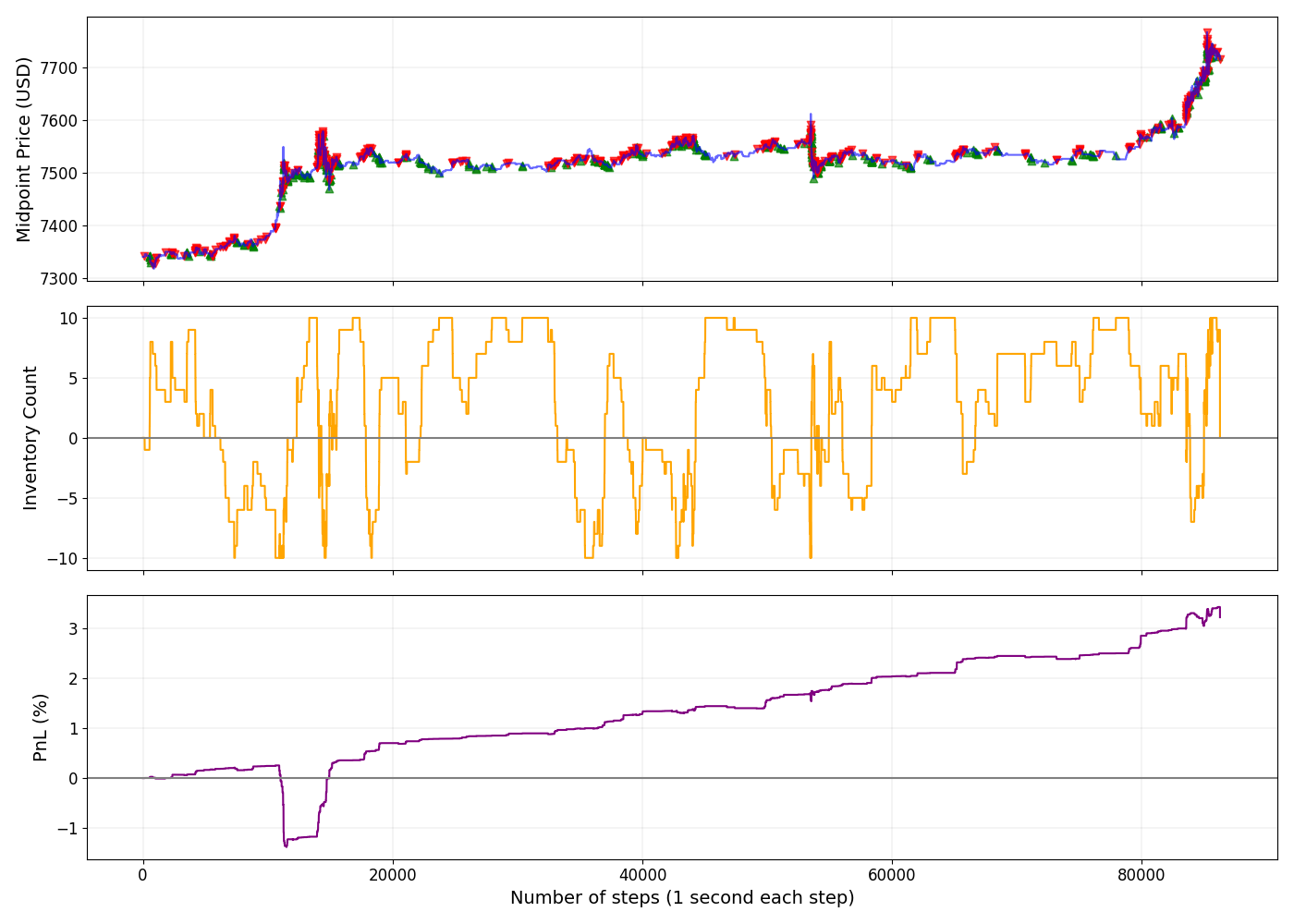}
            \hspace*{\fill}
            \caption{Plots of agent episode performance.   
            \textit{Green} and \textit{red} dots represent buy and sell executions, respectively.
            \textit{Left:} Example of price-based PPO agent making nearsighted decisions and frequent use of market orders with reward function $UPnL$ on February 3, 2020.  
            \textit{Right:} Example of time-based PPO agent trading tactically while failing to exploit price jumps with reward function $AsymC$ on January 6, 2020.}
            \label{fig:weakRewardFunctions}
        \end{figure}

        \begin{figure}[!ht]
            \hfill 
            \includegraphics[scale=.22]{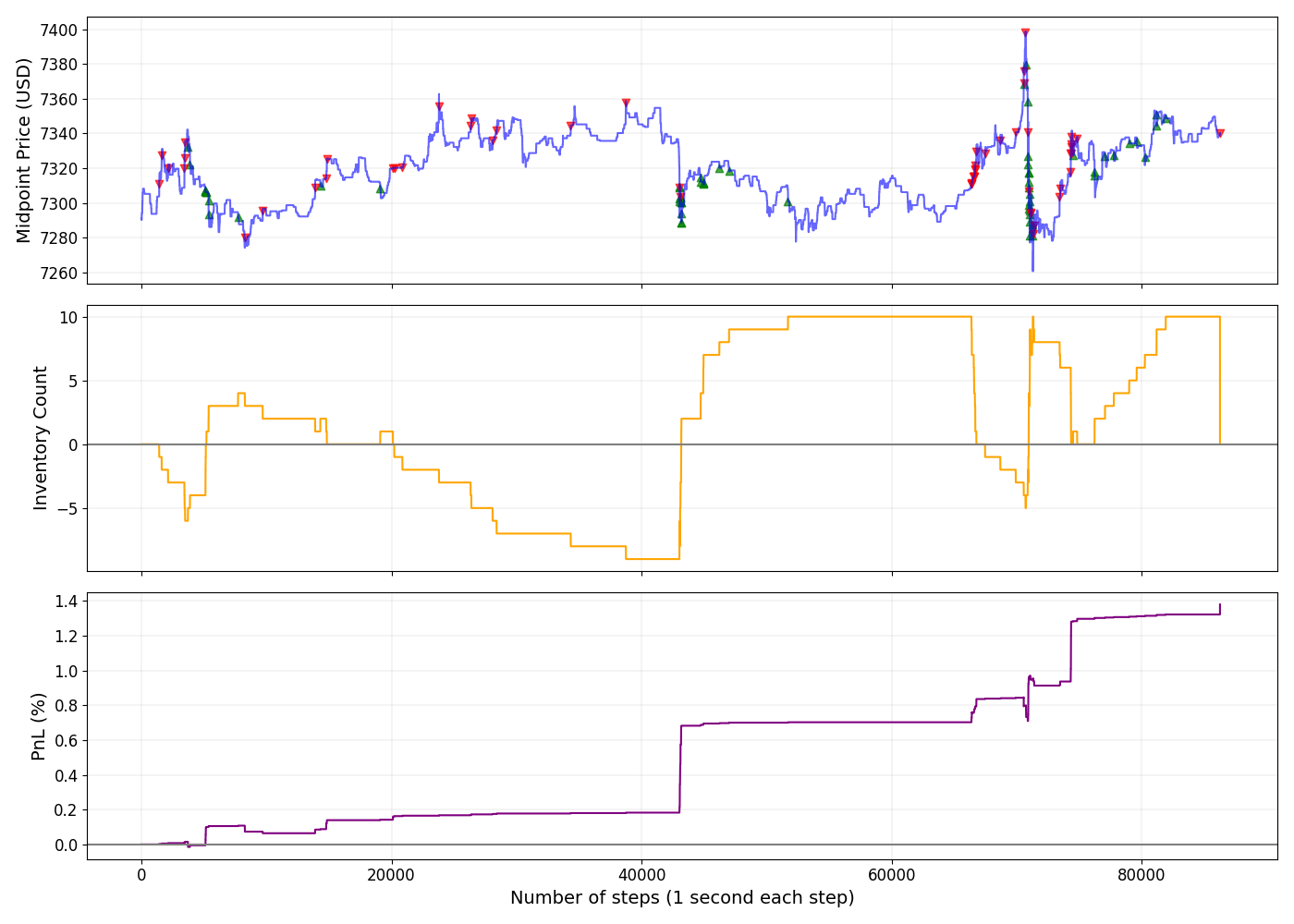}
            \includegraphics[scale=.22]{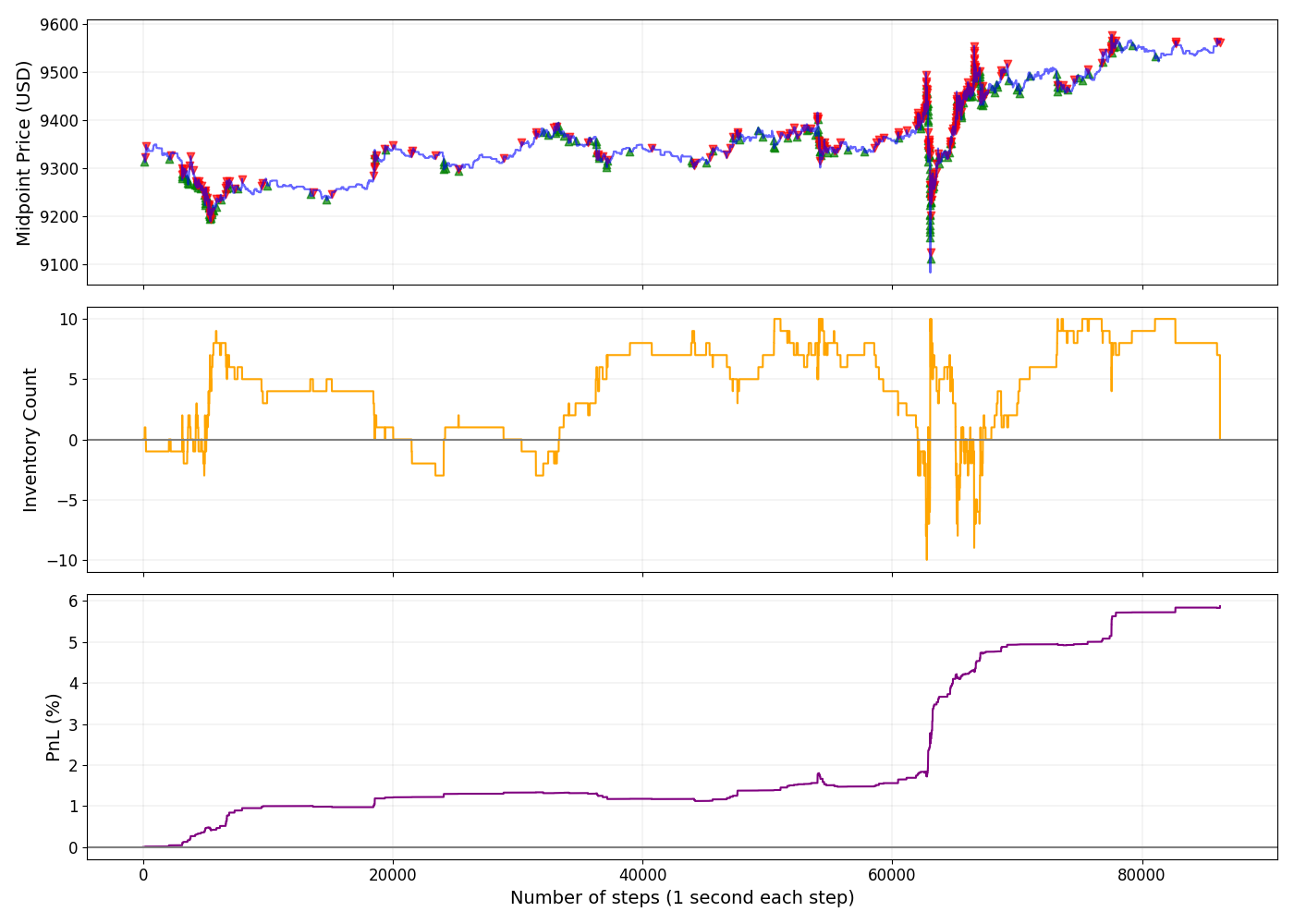}
            \hspace*{\fill}
            \caption{Plots of agent episode performance.   
            \textit{Green} and \textit{red} dots represent buy and sell executions, respectively.
            \textit{Left:} Example of time-based A2C agent effectively scaling into positions with goal-oriented reward function $TC$ on January 4, 2020.
            \textit{Right:} Example of price-based A2C agent actively trading and exploiting a price jump with reward function $DSR$ on January 30, 2020.}
            \label{fig:strongRewardFunctions}
        \end{figure}

    \subsection{Time-based Events} \label{Results:TimeEvents}
        The time-based environments were more difficult for the agents to learn; 15 out of 84 experiments led to profitable outcomes.  This is likely due to the training methodology, where agents may benefit from training for more than one million steps.  That said, the time-based environment was able to achieve the highest return out of all experiments due to quicker reactions to adverse price movements with the goal-based reward function $TC$.
    
        \begin{table}
            \centering
            \resizebox{\textwidth}{!}{
                \begin{tabular}{c*{12}{c}}
        			\toprule\toprule
                    \multicolumn{1}{c}{\textbf{Time-event:}} & \multicolumn{12}{c}{\textbf{Profit-and-Loss (\%)\footnotemark[\value{footnote}] }} \\ \hline
                    \multicolumn{1}{c}{\textbf{}} & \multicolumn{6}{c|}{\textbf{A2C}} & \multicolumn{6}{c}{\textbf{PPO}} \\ \hline
                    \multicolumn{1}{l|}{\textit{\textbf{Reward Function}}} & \multicolumn{1}{c}{\textit{\textbf{Set 1}}} & \multicolumn{1}{c}{\textit{\textbf{Set 2}}} & \multicolumn{1}{c}{\textit{\textbf{Set 3}}} & \multicolumn{1}{c}{\textit{\textbf{Set 4}}} & \multicolumn{1}{c}{\textit{\textbf{Set 5}}} & \multicolumn{1}{c|}{\textit{\textbf{Set 6}}} & \multicolumn{1}{c}{\textit{\textbf{Set 1}}} & \multicolumn{1}{c}{\textit{\textbf{Set 2}}} & \multicolumn{1}{c}{\textit{\textbf{Set 3}}} & \multicolumn{1}{c}{\textit{\textbf{Set 4}}} & \multicolumn{1}{c}{\textit{\textbf{Set 5}}} & \multicolumn{1}{c}{\textit{\textbf{Set 6}}} \\ \hline
                    \multicolumn{1}{l|}{\textit{UPnL}} & (-12.05) & (-11.67) & (-12.06) & (-24.00) & (-35.30) & (-14.83) & \multicolumn{1}{|c}{(-18.58)} & (-29.57) & (-25.62) & (-43.95) & (-32.12) & (-56.52) \\ \hline
                    \multicolumn{1}{l|}{\textit{UPnLwF}} & 4.53 & (-32.97) & 11.04 & (-4.73) & 5.12 & (-22.56) & \multicolumn{1}{|c}{(-5.44)} & (-20.34) & 1.56 & (-28.20) & (-17.37) & (-12.68) \\ \hline
                    \multicolumn{1}{l|}{\textit{Asym}} & (-13.09) & (-39.02) & (-35.41) & (-6.96) & 2.82 & 8.28 & \multicolumn{1}{|c}{(-14.00)} & (-16.61) & (-13.00) & (-17.97) & (-11.57) & (-4.69) \\ \hline
                    \multicolumn{1}{l|}{\textit{AsymC}} & 9.30 & 5.78 & 3.91 & 1.24 & (-15.92) & (-24.71) & \multicolumn{1}{|c}{(-18.04)} & (-15.49) & (-20.5) & \textbf{2.13} & (-13.32) & (-38.37) \\ \hline
                    \multicolumn{1}{l|}{\textit{$\Delta$ RPnL}} & (-10.57) & (-36.18) & (-19.41) & (-21.71) & (-33.18) & (-26.42) & \multicolumn{1}{|c}{(-31.16)} & (-39.70) & (-13.90) & (-8.19) & (-19.50) & (-30.56) \\ \hline
                    \multicolumn{1}{l|}{\textit{TC}} & (-7.22) & (-32.49) & \textbf{17.61} & (-7.83) & (-0.82) & 3.45 & \multicolumn{1}{|c}{(-24.88)} & (-2.42) & (-18.55) & (-13.28) & (-24.53) & (-19.34) \\ \hline
                    \multicolumn{1}{l|}{\textit{DSR}} & (-16.55) & (-23.33) & (-0.66) & 0.18 & 9.98 & (-2.99) & \multicolumn{1}{|c}{(-7.55)} & (-18.71) & (-6.11) & (-19.96) & (-28.43) & (-38.10)  \\  \bottomrule
                \end{tabular}
            }
    		\caption{Total return (in percentage) for out-of-sample data set (January 4, 2020 to March 3, 2020) using the \textit{time}-based event environment.}
    		\label{table:TimeEvent}
        \end{table}

    \subsection{Price-based Events}
        The price-based environments were easier to learn for the agents; 23 out of 84 experiments led to profitable outcomes.  This is likely due to the nature of having the agent take steps in the environment only when the price changes, therefore avoiding some noise in the LOB data.  Although this environment approach did not yield the highest score, in general the agent trading patterns appeared to be more stable and less erratic during large price jumps.
        
        \begin{table}[h t]
            \centering
            \resizebox{\textwidth}{!}{
                \begin{tabular}{c*{12}{c}}
        			\toprule\toprule
                    \multicolumn{1}{c}{\textbf{Price-event:}} & \multicolumn{12}{c}{\textbf{Profit-and-Loss (\%)\footnotemark[\value{footnote}] }} \\ \hline
                    \multicolumn{1}{c}{\textbf{}} & \multicolumn{6}{c|}{\textbf{A2C}} & \multicolumn{6}{c}{\textbf{PPO}} \\ \hline
                    \multicolumn{1}{l|}{\textit{\textbf{Reward Function}}} & \multicolumn{1}{c}{\textit{\textbf{Set 1}}} & \multicolumn{1}{c}{\textit{\textbf{Set 2}}} & \multicolumn{1}{c}{\textit{\textbf{Set 3}}} & \multicolumn{1}{c}{\textit{\textbf{Set 4}}} & \multicolumn{1}{c}{\textit{\textbf{Set 5}}} & \multicolumn{1}{c|}{\textit{\textbf{Set 6}}} & \multicolumn{1}{c}{\textit{\textbf{Set 1}}} & \multicolumn{1}{c}{\textit{\textbf{Set 2}}} & \multicolumn{1}{c}{\textit{\textbf{Set 3}}} & \multicolumn{1}{c}{\textit{\textbf{Set 4}}} & \multicolumn{1}{c}{\textit{\textbf{Set 5}}} & \multicolumn{1}{l}{\textit{\textbf{Set 6}}} \\ \hline
                    \multicolumn{1}{l|}{\textit{UPnL}} & (-31.42) & (-28.65) & (-38.74) & (-0.95) & (-0.91) & (-43.58) & \multicolumn{1}{|c}{(-31.74)} & (-25.89) & (-21.37) & (-46.12) & (-16.72) & (-32.32) \\ \hline
                    \multicolumn{1}{l|}{\textit{UPnLwF}} & 8.80 & (-11.85) & (-18.37) & (-21.66) & (-8.98) & (-20.24) & \multicolumn{1}{|c}{(-4.61)} & (-23.18) & 3.92 & (-9.92) & 3.50 & (-23.28) \\ \hline
                    \multicolumn{1}{l|}{\textit{Asym}} & (-27.21) & (-2.00) & (-1.66) & (-0.86) & (-14.82) & (-8.16) & \multicolumn{1}{|c}{(-16.04)} & (-12.76) & (-15.32) & (-6.10) & (-15.78) & (-10.73) \\ \hline
                    \multicolumn{1}{l|}{\textit{AsymC}} & (-7.12) & (-11.58) & (-12.24) & 11.88 & (-14.98) & (-14.12) & \multicolumn{1}{|c}{(-19.58)} & (-15.92) & (-2.08) & (-12.57) & (-8.21) & (-15.42) \\ \hline
                    \multicolumn{1}{l|}{\textit{$\Delta$ RPnL}} & 2.65 & (-1.62) & 6.70 & 8.74 & 6.97 & 6.71 & \multicolumn{1}{|c}{(-6.29)} & (-6.28) & (-13.76) & 3.60 & (-11.47) & (-28.35) \\ \hline
                    \multicolumn{1}{l|}{\textit{TC}} & 6.28 & 10.66 & 10.19 & 6.91 & \textbf{13.43} & 9.57 & \multicolumn{1}{|c}{5.72} & (-23.71) & (-29.99) & 2.73 & \textbf{11.38} & (-16.25) \\ \hline
                    \multicolumn{1}{l|}{\textit{DSR}} & (-27.35) & 2.80 & 12.23 & (-1.80) & 9.73 & (-28.14) & \multicolumn{1}{|c}{(-14.23)} & (-13.96) & (-19.67) & (-24.19) & 5.74 & (-33.25) \\ \bottomrule
                \end{tabular}
            }
    		\caption{Total return (in percentage) for out-of-sample data set (January 4, 2020 to March 3, 2020) using the \textit{price}-based event environment.}
    		\label{table:PriceEvent}
        \end{table}

\section{Conclusion} \label{Conclusion}
    In this paper, two advanced policy-based model-free reinforcement learning algorithms were trained to learn automated market making for Bitcoin using high resolution Level II tick data from Bitmex exchange. The agents learned different trading strategies from seven different reward functions and six different combinations of features for the agent’s observation space. Additionally, this paper proposes a price-based approach to defining an event in which the agent steps through the environment and demonstrates its effectiveness to solve the automated market making challenge, extending the DRLMM framework \cite{sadighian2019deep}. 

    All agents were trained for one million steps across eight days of data and evaluated on 30 out-of-sample days.  The A2C algorithm outperformed PPO in terms of cumulative return and number of profitable experiments.  An A2C agent with goal-based $TC$ reward function generated the greatest return for both time- and price-based environments.
    
    Several observations made during the execution of this experiment could lead to fruitful future research avenues.  First, a formalized methodology for training model-free reinforcement learning in context of financial time-series problem.  More specifically, it would be worthwhile to explore the effects of a framework for scoring and selecting which trading days to include in the training data set (e.g., volatility, daily volume, number of price jumps, etc.).  Second, with the demonstrated success of more advanced neural network architectures in the supervised learning domain \cite{Zhang_2019, wallbridge2020transformers}, it would be interesting to see if convolution, attention, and recurrent neural networks help the agents learn to better exploit price jumps.

\paragraph{ACKNOWLEDGEMENTS}
    Thank you to Toussaint Behaghel for reviewing the paper and providing helpful feedback and Florian Labat for suggesting the use of price-based events in reinforcement learning.  Thank you to Mathieu Beucher, Sakina Ouisrani, and Badr Ghazlane for helping execute experiments and collate results.
    
\bibliography{references}
\bibliographystyle{unsrt}
\newpage

\appendix

\section{Appendix} \label{Appendix}

    \subsection{Agent Configurations} \label{AgentConfigurations}
	The parameters used to train agents in all experiments.
	\begin{table}[h	t]
		\centering 
		\begin{tabular}{p{0.02\textwidth} p{0.2\textwidth} p{0.34\textwidth}}
			\toprule\toprule
			\textbf{\#} & \textbf{Parameter} & \textbf{Value} \\ [0.5ex] 
			\hline
			1 & Action repeats & 5 \\
			\hline
			2 & Window size & 100 \\
			\hline
			3 & Transaction fees & Limit -0.025\% / Market 0.075\%  \\
			\hline
			4 & Max positions & 10 \\
			\hline
			5 & Gamma $\gamma$ & 0.99 \\
			\hline
			6 & Learning rate $\alpha$ & 3e-4 \\
			\hline
			7 & No. of LOB levels & 20 \\
			\hline
			8 & K-steps & 256 (PPO) / 40 (A2C) \\
			\hline
			9 & Training steps & 1,000,000 \\
			\hline
			10 & Action space & 17 \\
			\hline
			11 & Dampening & 0.35 \\
			\hline
			12 & GAE $\lambda$ & 0.97 \\
			\hline
			13 & Price-event threshold & 0.01\% \\
			\hline
			14 & Optimizer & Adam (both A2C and PPO) \\
			\bottomrule
		\end{tabular}
		\caption{Parameters for experiments.}
		\label{table:hyperparameters}
	\end{table}

    \subsection{Observation Space} \label{AgentObservationSpace}
    	As set forth in our previous work \cite{sadighian2019deep}, the agent’s observation space is a combination of three sub-spaces: the environment state space, consisting of LOB, trade and order flow snapshots with a window size $w$; the agent state space, consisting of handcrafted risk and position indicators; and the agent action space, consisting of a one-hot vector of the agent’s latest action.  In this experiment, $w$ is set to 100.
    
        \subsubsection{Environment State Space} \label{ess}
        	
        	\paragraph{LOB Quantity}
            	The dollar value of each price level in the LOB, where $\chi$ is the dollar value at LOB level $i$ at time $t$, applied to both $bid$ and $ask$ sides. Since we use the first 20 price levels of the LOB, this feature is represented by a vector of 40 values.
            	
            	\begin{equation}
            		\chi^{bid, ask}_{t, i} =\sum_{i=0}^{I-1} p_{t, i}^{bid, ask} \times q_{t, i}^{bid, ask}
            	\end{equation}
            	where $p^{bid, ask}$ is the price and $q^{bid, ask}$ is the quantity at LOB level $i$ for bid and ask sides, respectively.
        	
        	\paragraph{LOB Imbalances}
            	The order imbalances $\iota \in [-1, 1]$ are represented by the cumulative dollar value for each price level $i$ in the LOB.  Since we use the first 20 price levels of the LOB, this feature is represented by a vector of 20 values.
            	
            	\begin{equation}
            		\iota_{t, i} = \frac{\chi^{ask, q}_{t, i} - \chi^{bid, q}_{t, i}}{\chi^{ask, q}_{t, i} + \chi^{bid, q}_{t, i}}
            	\end{equation}
        	
        	\paragraph{Order Flow} \label{OrderFlowStatistics}
            	The sum of dollar values for cancel $C$, limit $L$, and market $M$ orders is captured between each LOB snapshot.  Since we use the first 20 price levels of the LOB, this feature is represented by a vector of 120 values, 60 per each side of the LOB.
            	
            	\begin{equation}
            		C_{t, i}^{bid, ask} = p_{t, i}^{bid, ask} \times q_{t, i}^{bid, ask}
            	\end{equation}
            	\begin{equation}
            		L_{t, i}^{bid, ask} = p_{t, i}^{bid, ask} \times q_{t, i}^{bid, ask}
            	\end{equation}
            	\begin{equation}
            		M_{t, i}^{bid, ask} = p_{t, i}^{bid, ask} \times q_{t, i}^{bid, ask}
            	\end{equation}
            	where $q$ is the number of units available at price $p$ at LOB level $i$.
        	
        	\paragraph{Trade Flow Imbalances}
            	The Trade Flow Imbalances $TFI \in [-1, 1]$ indicator measures the magnitude of buyer initiated $BI$ and and seller initiated $SI$ transactions over a given window $w$. Since we use 3 different windows $w$ (5, 15, and 30 minutes), this feature is represented by a vector of 3 values.
            	
            	\begin{equation} \label{eq:tfi}
            		TFI_{t} = \frac{UP_{t} - DWN_{t}}{UP_{t} + DWN_{t}}
            	\end{equation}
            	where $UP_{t} = \sum_{n=0}^{w} BI_{n}$ and $DWN_{t} = \sum_{n=0}^{w} SI_{n}$.
        	
        	\paragraph{Custom RSI}
            	The relative strength index indicator (RSI) measures the magnitude of prices changes over a given window $w$. This custom implementation $CRSI \in [-1, 1]$ scales the data so that it does not require normalization, even though we do use the scaled z-score values in our experiment.  Since we use 3 different windows $w$ (5, 15, and 30 minutes), this feature is represented by a vector of 3 values.
            	
            	\begin{equation}
            	    CRSI_{t} = \frac{gain_{t} - \arrowvert loss_{t} \arrowvert}{gain_{t} + \arrowvert loss_{t} \arrowvert}
            	\end{equation}	
            	where	
            	$gain_{t} = \sum_{n=0}^{w} \Delta m_{n} \textnormal{ if } \Delta m_{n} > 0 \textnormal{ else } 0$
            	and	
            	$loss_{t} = \sum_{n=0}^{w} \Delta m_{n} \textnormal{ if } \Delta m_{n} < 0 \textnormal{ else } 0$
            	and	
            	$\Delta m_{t} = \frac{m_{t}}{m_{t-1}} - 1$.
        	
        	\paragraph{Spread}
            	The spread $\varsigma_{t}$ is the difference between the best bid $p^{bid}$ and best ask $p^{ask}$.  This feature is represented as a scalar.
            
            	\begin{equation}
            		\varsigma_{t} = p_{t}^{bid} - p_{t}^{ask}
            	\end{equation}
        	
        	\paragraph{Change in Midpoint}
            	The change in midpoint $\delta m_t$ is the log difference in midpoint prices between time step $t$ and $t-1$.  This feature is represented as a scalar.
            
            	\begin{equation}
            		\delta m_{t} = \log m_{t} - \log m_{t-1}
            	\end{equation}
            	
        	\paragraph{Reward}
        	    The reward $r$ from the environment, as described in section \ref{RewardFuctions}.

    	\subsubsection{Agent State Space}
        	
        	\paragraph{Net Inventory Ratio}
            	The agent’s net inventory ratio $\upsilon \in [-1, 1]$ is the inventory count $Inv$ represented as a percentage of the maximum inventory $IM$.  This feature is represented as a scalar.
            	
            	\begin{equation}
            		\upsilon_{t} = \frac{Inv^{long} - Inv^{short}}{IM}
            	\end{equation}
        	
        	\paragraph{Realized PnL}
            	The agent’s realized profit-and-loss $RPnL$ is the sum of realized and unrealized profit and losses.  In this experiment, the $RPnL$ is scaled by a scalar value $\rho$, which represents the daily PnL target.
        	
        	\paragraph{Unrealized PnL}
            	The agent’s current unrealized PnL $UPnL_t$ is the unrealized PnL across all open positions.  The unrealized PnL feature is represented as a scalar, containing the net of $\textnormal{long}$ and $\textnormal{short}$ positions.
            	
            	\begin{equation}
            		UPnL_t = \left[ \frac{p_{t}^{Avg, short}}{m_t} - 1\right] + \left[ \frac{m_t}{p_{t}^{Avg, long}} - 1 \right]
            	\end{equation}
            	where $p_{Avg}$ is the average price of the agent's $long$ or $short$ position and $m$ is the midpoint price at time $t$.
        	
        	\paragraph{Open Order Distance to Midpoint}
            	The agent’s open limit order distance to midpoint is the distance $\zeta$ of the agent’s open $bid$ and $ask$ limit orders $L$ to the midpoint price $m$ at time $t$. The feature is represented as a vector with 2 values.
            	
            	\begin{equation}
            		\zeta^{long, short}_{t} = \frac{L^{bid, ask}_{t}}{m_t} - 1
            	\end{equation}
        	
        	\paragraph{Order Completion Ratio}
            	Order completion $\eta \in [-1, 1]$ is a custom indicator that incorporates an order’s relative position in the LOB queue $\kappa$ and partial executions $Ex$ relative to the order size $Sz$. The feature is represented as a vector with 2 value, one per $long$ and $short$ sides.
            	
            	\begin{equation}
            		\eta_{t}^{long, short} = \frac{Ex_{t}^{long, short} - \kappa_{t}^{long, short}}{\kappa_{t}^{long, short} + Sz_{t}^{long, short}}
            	\end{equation}
        	
        	\paragraph{Agent Action Space}
            	The agent's action space is included in the agent state space.  It is represented by a one-hot vector over the action space outlined in table \ref{table:action_space}.

\end{document}